\documentclass[12pt]{article}
\usepackage{titlesec}
\usepackage{geometry}
\usepackage{fancyhdr}
\usepackage{xcolor}
\usepackage{amsfonts}
\usepackage[T1]{fontenc}
\usepackage{listings} 
\usepackage{xcolor}   
\usepackage{graphicx}
\usepackage{parskip} 
\usepackage{amsmath}
\usepackage{lipsum} 
\usepackage{changepage} 
\usepackage{tikz} 
\usepackage{newpxtext, newpxmath} 
\usepackage{mdframed} 
\usepackage{tocloft} 
\usepackage{float} 

\lstdefinestyle{mystyle}{
    language=Python,
    basicstyle=\ttfamily\small,
    backgroundcolor=\color{white},
    commentstyle=\color{green},
    keywordstyle=\color{blue},
    stringstyle=\color{red},
    showstringspaces=false,
    breaklines=true,
    frame=single,
    numbers=left,
    numberstyle=\tiny\color{gray},
    captionpos=b,
    tabsize=4
}

\fancypagestyle{plain}{ 
  \fancyhf{} 
  \fancyfoot[C]{\color{gray}\small\thepage} 
}
\titleformat{\section}
  {\normalfont\Large\bfseries\scshape\centering\color{blue}} 
  {\thesection}{1em}{}
\titlespacing*{\section}{0pt}{1.5ex plus 0.5ex minus 0.2ex}{1.5ex plus 0.5ex minus 0.2ex} 

\titleformat{\subsection}
  {\normalfont\large\bfseries\scshape\color{blue}} 
  {\thesubsection}{1em}{}
\titlespacing*{\subsection}{0pt}{1.2ex plus 0.5ex minus 0.2ex}{1.2ex plus 0.5ex minus 0.2ex} 

\usepackage[colorlinks=true]{hyperref}
\hypersetup{
  citecolor=green, 
  linkcolor=red,  
}

\begin{document}

\begin{titlepage}
  \begin{center}
    {\huge\bfseries Hawkes Processes in High-Frequency Trading} \\[1cm]
    {\large Modeling High-Frequency Data with Bivariate Hawkes Processes: Power-Law vs. Exponential Kernels} \\[1cm] 
    {\normalsize Neal Batra} \\[0.1cm]
    {\footnotesize\itshape Mathematical Finance, Warwick Finance Societies \\ 
E-mail: neal.batra@warwick.ac.uk} \\[1cm]
  \end{center}
\end{titlepage}

\begin{adjustwidth}{0.25in}{0.25in} 
\vspace*{0.1em} 
\noindent\rule{\dimexpr\textwidth-0.5in\relax}{0.5pt} 
\vspace{0.5em} 
\begin{center}
  \normalfont\scshape\bfseries Abstract 
\end{center}
\vspace{0.5em} 
\noindent\normalfont 
This study explores the application of Hawkes processes to model high-frequency data in the context of limit order books. Two distinct Hawkes-based models are proposed and analyzed: one utilizing exponential kernels and the other employing power-law kernels. These models are implemented within a bivariate framework. The performance of each model is evaluated using high-frequency trading data, with a focus on their ability to reproduce key statistical properties of limit order books. Through a comprehensive comparison, we identify the strengths and limitations of each kernel type, providing insights into their suitability for modeling high-frequency financial data. Simulations are conducted to validate the models, and the results are interpreted. Based on these insights, a trading strategy is formulated.  

\vspace{0.5em} 
\noindent\rule{\dimexpr\textwidth-0.5in\relax}{0.4pt} 
\vspace{0.5cm} 
\end{adjustwidth}

\section*{Introduction}
High-frequency trading (HFT) has become a dominant force in modern financial markets, characterized by rapid order execution, low latency, and large volumes of data. At the heart of HFT lies the limit order book (LOB), a system that records all buy and sell orders for a given financial instrument at various price levels \cite{optiver}. Understanding the complex mechanisms of LOBs is crucial for market participants, as it provides insights into price formation, market liquidity, and the impact of order flow on asset prices.
\\~\\
To model the behavior of LOBs, Hawkes processes have emerged as a powerful mathematical tool. Hawkes processes are self-exciting point processes that capture the clustering of events, such as order arrivals, which are key in high-frequency data. In this study, we begin by introducing the one-dimensional Hawkes process, which models a single stream of events (e.g., buy orders or sell orders). We provide the necessary definitions and key concepts, such as the intensity function, kernels, and the role of self-excitation in capturing temporal dependencies.
\\~\\
Building on the one-dimensional framework, we extend our analysis to the multivariate case, where multiple interacting event streams are modeled simultaneously. This extension is essential for capturing the cross-exciting dynamics between different types of orders (e.g., buy and sell orders) in a limit order book. However, to maintain a balance between simplicity and realism, we focus on the bivariate Hawkes process, which considers two interacting event streams. This choice allows us to model the work between buy and sell orders while keeping the model interpretable.
\newpage
The primary objective of this study is to compare two types of kernels commonly used in Hawkes processes: Exponential kernels and Power-Law kernels (however there also exists the guassian kernel). Exponential kernels are widely used due to their mathematical simplicity and interpretability, while power-law kernels offer a more flexible framework for capturing long-memory effects and heavy-tailed behavior in high-frequency data. By implementing and comparing these kernels within a bivariate Hawkes process framework, we aim to evaluate their performance in reproducing key statistical properties of LOBs, such as order arrival rates, price impact, and clustering behavior.

\subsection*{Limit Order Books}
The limit order book (LOB) serves as a centralized database that records all outstanding buy and sell orders for a specific security on an exchange \cite{optiver}. The smallest price increment by which a security can move is referred to as a \textit{tick}. At any given time t, the highest price at which there is an outstanding buy order is known as the \textit{bid price}, while the lowest price for an outstanding sell order is called the \textit{ask price}. The difference between the ask and bid prices is defined as the \textit{bid-ask spread}, and the \textit{mid price} is calculated as the arithmetic average of the bid and ask prices. 
\section*{Mathematical Formulations}
\subsection*{Counting Processes}
\normalsize

In order to sufficiently introduce Hawkes Processes, an introduction to counting processes is required.

\textbf{Definition 1 (Counting Process)} A \textit{counting process} is a stochastic process \\ \((N(t) : t \geq 0)\) taking values in \(\mathbb{N}_0\) that satisfies the following properties:  
\begin{itemize}
    \item \(N(0) = 0\),
    \item \(N(t)\) is almost surely (a.s.) finite for all \(t \geq 0\),
    \item \(N(t)\) is a right-continuous step function with increments of size \(+1\).
\end{itemize}

Let \((\mathcal{F}(t) : t \geq 0)\) denote the history of arrivals up to time \(t\), where \(\mathcal{F}(\cdot)\) is a filtration, i.e., an increasing sequence of \(\sigma\)-algebras.

\textbf{Definition 2 (Conditional intensity function)} Let \((N(t) : t \geq 0)\) be a counting process adapted to a filtration \((\mathcal{F}(t) : t \geq 0)\), where \(\mathcal{F}(t)\) represents the history of the process up to time \(t\). The \textit{conditional intensity function} is defined as:  
\[
\lambda(t) = \lim_{\Delta t \to 0} \frac{\mathbb{E}\left[N(t + \Delta t) - N(t) \mid \mathcal{F}(t)\right]}{\Delta t}
\]  

provided the limit exists almost surely. 
Here:  
\begin{itemize}
    \item \(N(t + \Delta t) - N(t)\) counts the number of events in the interval \((t, t + \Delta t]\),
    \item \(\mathbb{E}\left[N(t + \Delta t) - N(t) \mid \mathcal{F}(t)\right]\) is the expected number of events in \((t, t + \Delta t]\) given the history up to time \(t\). 
\end{itemize}

The function \(\lambda(t)\) describes the instantaneous rate of events at time \(t\), conditioned on the history of the process up to time \(t\). Moreover, \( \lambda \) uniquely determines the distribution of the intervals between events. 

\subsection*{Hawkes Processes}

A Hawkes process \((N(t) : t \geq 0)\) in one dimension is a counting process defined by its intensity function \( \lambda(t) \), which represents the rate of event arrivals. The intensity function is given by \cite{oxford}:

\begin{equation}
\lambda_t = \mu + \sum_{t_i < t} \phi(t - t_i) = \mu + \int_{0}^t \phi(t-s) \, dN_s    
\end{equation}

where:
\begin{itemize}
    \item \( \mu > 0 \) is the base intensity (which represents the rate of events in the absence of past events)
    \item \( \phi \) is the excitation kernel that models the influence of past events \( t_i \) on the current intensity.
\end{itemize}

Now that the \textbf{Hawkes process for one dimension} has been introduced, we can explore how the \textbf{excitation kernel} \( \phi(t - t_i) \) determines the influence of past events on the current intensity. The choice of kernel function is crucial, as it defines the nature of the temporal dependencies in the process. Common forms of the kernel include \cite{mura}:
\\
\subsection*{Exponential Kernel}
The exponential kernel is defined as:
\begin{equation}
    \phi(t - t_i) = \alpha e^{-\beta (t - t_i)}
    \label{exponential}
\end{equation}

where:
\begin{itemize}
    \item \( \alpha > 0 \) controls the \textbf{strength} of the excitation,
    \item \( \beta > 0 \) controls the \textbf{decay rate} of the excitation.
\end{itemize}

\subsection*{Power-Law Kernel}
The power-law kernel is defined as:
\begin{equation}
\phi(t - t_i) = \frac{\alpha}{(t - t_i + \varepsilon)^\beta}
\label{powerlaw}    
\end{equation}
where:
\begin{itemize}
    \item \( \alpha > 0 \) controls the \textbf{strength} of the excitation,
    \item \( \beta > 0 \) controls the \textbf{decay rate} of the excitation,
    \item \( \varepsilon > 0 \) is a small constant added to avoid division by 0 at \( t = t_i \).
\end{itemize}

\subsection*{Transition To Bivariate Framework}
As explained in the introduction, our focus extends beyond the one-dimensional Hawkes process to a \textbf{bivariate framework}, which allows us to model interactions between two types of events (e.g., buy and sell orders in a limit order book). To generalize the intensity function to multiple dimensions, we define the \textbf{multivariate intensity function}. Consider a d-dimensional Hawkes Process. For each \( i,j \in \{1, 2, \dots, d\} \), the intensity function is given by:

\begin{equation}
\lambda_i(t) = \mu_i + \sum_{j=1}^d \int_{0}^t \phi_{ij}(t-s) \, dN_j(s)
\label{intensity}
\end{equation}
where:
\begin{itemize}
    \item \( \mu_i \) is the base intensity for the \( i \)-th component,
    \item \( \phi_{ij} \) is the excitation kernel capturing the influence of events from the \( j \)-th component on the \( i \)-th component,
    \item \( N_j(s) \) is the counting process for the \( j \)-th component.
\end{itemize}
For our purposes, we will restrict the framework to \textbf{two variables}, setting d = 2 results in a bivariate Hawkes process.
\begin{equation}
\lambda_i(t) = \mu_i + \sum_{j=1}^2 \int_{0}^t \phi_{ij}(t-s) \, dN_j(s), \quad  i,j \in \{1, 2\} \ 
\label{intensity}
\end{equation}
In particular, if the kernel \( \phi_{ij}(t - s) \) takes the form of an \textbf{exponential kernel} \eqref{exponential}:
\[
\phi_{ij}(t - s) = \alpha_{ij} e^{-\beta_{ij} (t - s)},
\]
then, from \eqref{intensity} the intensity function for the \( i \)-th component becomes:
\begin{equation}
\boxed{\lambda_i(t) = \mu_i + \sum_{j=1}^2 \int_0^t \frac{\alpha_{ij}}{e^{\beta_{ij} (t - s)}}  \, dN_j(s), \quad  i,j \in \{1, 2\} \ }
\label{ebiv}
\end{equation}

where:
\begin{itemize}
    \item \( \alpha_{ij} > 0 \) controls the strength of excitation from the \( j \)-th component to the \( i \)-th component,
    \item \( \beta_{ij} > 0 \) controls the decay rate of the excitation,
    \item The condition \( \frac{\alpha_{ij}}{\beta_{ij}} < 1 \) ensures stationarity of the process.
\end{itemize}

Similarly, if the kernel \( \phi_{ij}(t - s) \) takes the form of a \textbf{power-law kernel} \eqref{powerlaw}:
\[
\phi_{ij}(t - s) = \frac{\alpha_{ij}}{(t - s + \varepsilon_{ij})^{\beta_{ij}}},
\]
then, from \eqref{intensity} the intensity function for the \( i \)-th component becomes:
\begin{equation}
\boxed{\lambda_i(t) = \mu_i + \sum_{j=1}^2 \int_0^t \frac{\alpha_{ij}}{(t - s + \varepsilon_{ij})^{\beta_{ij}}} \, dN_j(s), \quad  i,j \in \{1, 2\} \ }   
\label{plawbiv}
\end{equation}

With the mathematical framework established and the intensity functions defined by \eqref{ebiv} and \eqref{plawbiv}, the next step is to validate and explore the model through a series of simulations. These simulations are designed to test the model's ability to replicate real-world market dynamics, compare its performance under different assumptions, and derive actionable insights. Specifically, we will conduct four types of simulations: \\ \textbf{(1) Fitting the model to empirical data to analyze observed event patterns, \\ (2) Generating synthetic data using the fitted parameters to validate the model's accuracy \\ (3) Comparing the empirical and simulated intensity functions to assess the model's robustness\\ (4) Testing an exponential kernel as an alternative to the power-law kernel to determine which better captures market behavior.} 

\section*{Simulation}

To validate the Hawkes model and explore its ability to replicate real-world market dynamics, we will conduct a series of simulations using high-frequency trading data obtained from Alpha Vantage and LOBSTER \cite{alpha}. The dataset will focus on AMD and TESLA stocks over a 2-second window on March 14th, 2025, from 10:23:00 to 10:23:02. The data will then be tediously processed by me into a structured csv file through which we will use:
\begin{itemize}
    \item (\texttt{pandas}) for loading, cleaning, and preprocessing the CSV data.

    \item (\texttt{numpy}) for numerical computations and handling arrays.

    \item (\texttt{matplotlib}) and (\texttt{scienceplots}) for creating high-quality visualizations of the intensity functions and simulation results.

    \item (\texttt{scipy.optimize}) for fitting the Hawkes process parameters using the L-BFGS-B optimization algorithm.
\end{itemize}

To estimate the parameters of the Hawkes process, we will minimize the \textbf{negative log-likelihood function} using the \textbf{L-BFGS-B} optimization algorithm. This function quantifies how well the model fits the observed data, and minimizing it will yield the optimal parameter values ($\mu_i$, $\alpha_{ij}$, $\beta_{ij}$). The \texttt{minimize} function will take the following inputs: the negative log-likelihood function, an initial guess for the parameters (\texttt{initial\_params}), and additional arguments such as the observed buy and sell event times (\texttt{buy\_times} and \texttt{sell\_times}). The \texttt{method} parameter will be set to \texttt{'L-BFGS-B'}, which is a highly efficient algorithm for solving bounded optimization problems. Additionally, bounds will be specified for each parameter to ensure they remain within physically meaningful ranges (e.g., $\alpha_{ij} > 0$, $\beta_{ij} > 0$). The result of this optimization will provide the fitted parameters, which will be used to compute the intensity functions and analyze the model's performance.

\subsection*{Power-Law Kernel Empirical \& Simulated Data Intensity Function}
After processing over 200 events from the empirical dataset, we computed the intensity functions for buy and sell events using the fitted parameters from the power-law kernel. Below are the plots of the intensity function against time for the real data and simulated data. Showcasing how well the Hawkes process model captures the observed market dynamics. The intensity functions reveal the temporal clustering of events, with buy and sell intensities exhibiting distinct patterns of self-excitation and cross-excitation. Specifically, the buy intensity function shows sharp peaks during periods of high market activity, indicating strong self-excitation, while the sell intensity function displays smoother trends with occasional spikes, reflecting weaker self-excitation and potential cross-excitation from buy events. 
\begin{figure}[h!]
    \centering
    \includegraphics[width=1\textwidth]{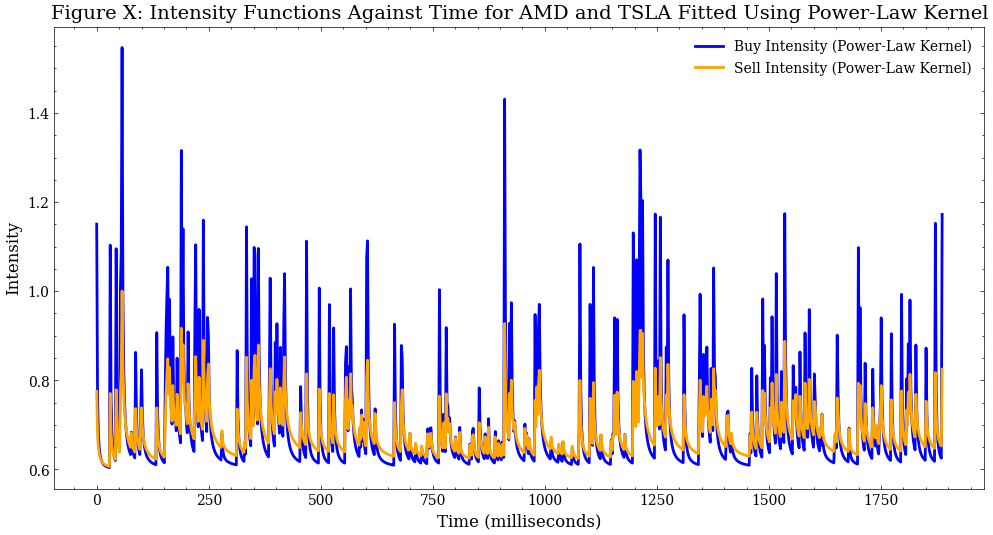} 
    \caption{\textit{Intensity Functions Against Time for Real Data Fitted Using the Power-Law Kernel. The buy intensity is shown in blue and the sell intensity in orange.}}
    \label{fig:intensity_real}
\end{figure}

The sell intensity (orange) appears to be a scaled-down version of the buy intensity (blue), suggesting that selling orders occur at a lower multiple of buy orders. This could indicate that \textbf{market participants or trading algorithms are more aggressively placing buy orders, possibly due to upward momentum or liquidity-seeking strategies.}

\textbf{The structured nature of the spikes strongly suggests algorithmic trading rather than human-driven activity}. Given that the data spans only \textit{2} seconds, the presence of clear and frequent peaks indicates automated execution. This could be due to market-making, where an algorithm continuously places and cancels orders.

Additionally, If a human trader were executing these orders, the timing and intensity fluctuations would likely appear more random. Instead, the highly structured nature of these patterns confirms the influence of high-frequency trading (HFT) strategies operating at millisecond precision.

\begin{figure}[H] 
    \centering
    \includegraphics[width=1\textwidth]{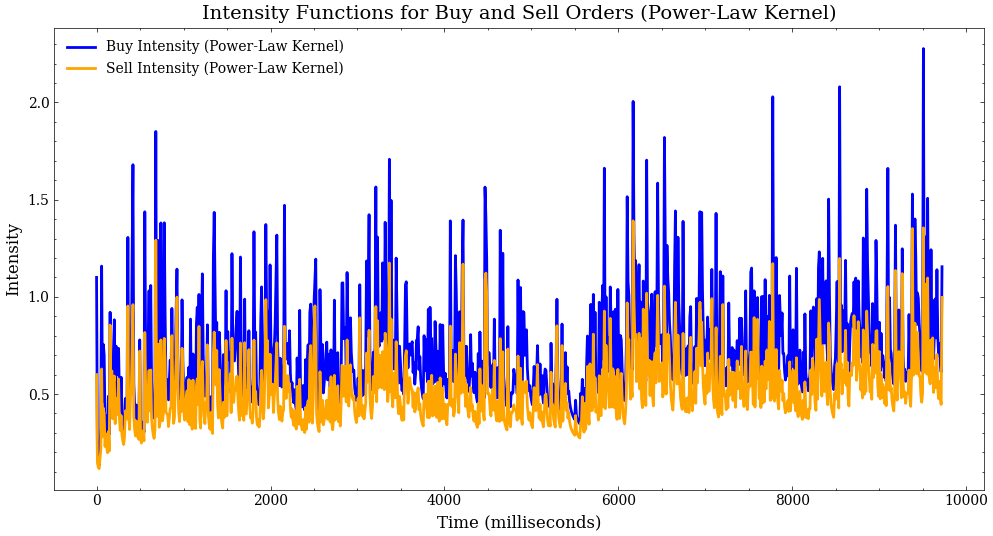} 
    \caption{\textit{Intensity Functions Against Time for Simulated Data Fitted Using the Power-Law Kernel. The buy intensity (blue) exhibits sharp peaks during periods of high activity, while the sell intensity (orange) displays smoother trends with occasional spikes.}}
    \label{fig:intensity_simulated}
\end{figure}

\subsection*{Exponential Kernel Empirical \& Simulated Data}

\begin{figure}[H] 
    \centering
    \includegraphics[width=1\textwidth]{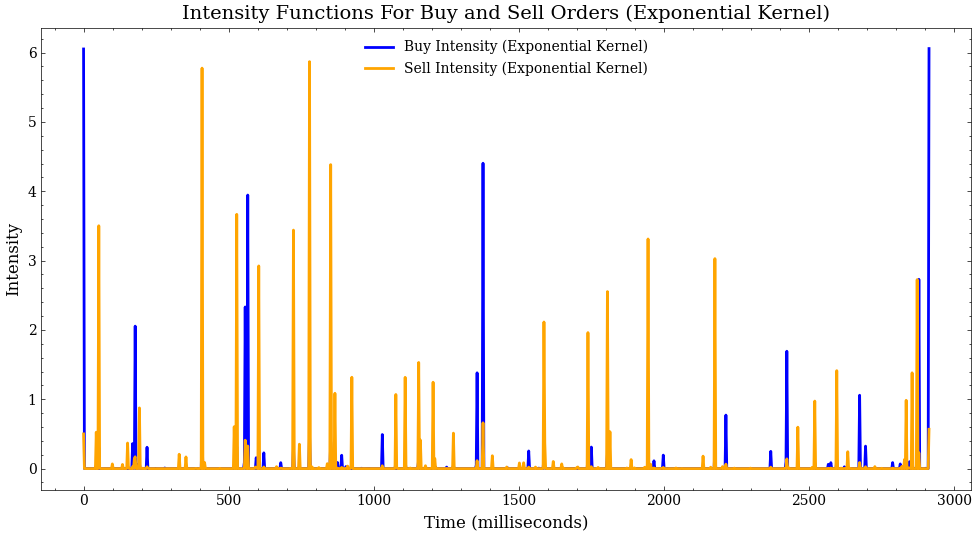} 
    \caption{\textit{Intensity Functions Against Time for Simulated Fitted Using the Exponential Kernel. The buy intensity (blue) exhibits fewer peaks compared to the sell intensity (yellow). This implies the exponential kernel has certain limitations.}}
    \label{fig:intensity_plot}
\end{figure}

\newpage

\begin{figure}[h!]
    \centering
    \includegraphics[width=1\textwidth]{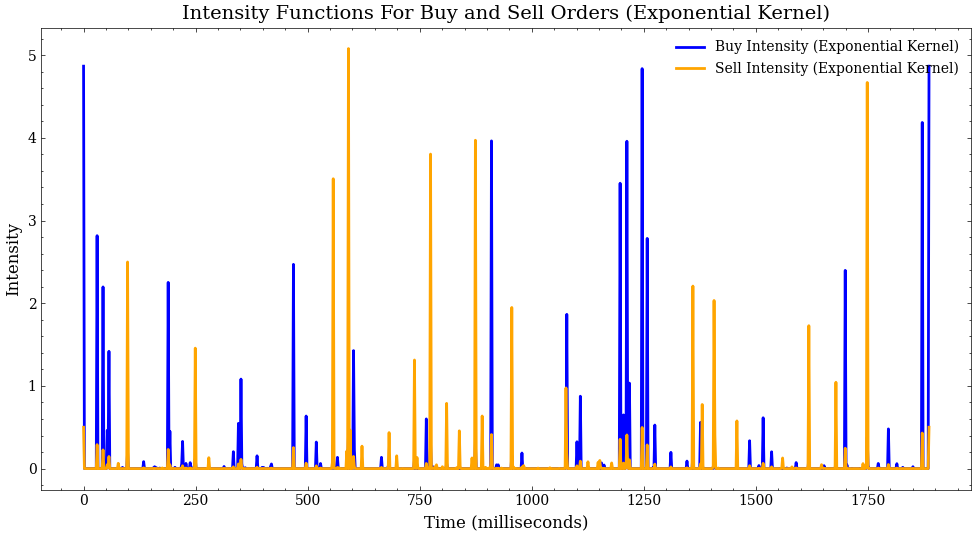} 
    \caption{\textit{Intensity Functions Against Time for Simulated Fitted Using the Exponential Kernel. The buy intensity (blue) exhibits less peaks compared to the sell intensity (yellow). Implies the exponential kernel has certain limitations.}}
    \label{fig:intensity_plot}
\end{figure}

\subsection*{Evaluation}
\textbf{The power-law kernel demonstrates a significantly better fit}, as evidenced by the sharp peaks in the buy intensity (blue) and the smoother yet responsive trends in the sell intensity (orange). These characteristics align well with the empirical behavior of HFT data, where trading activity often exhibits bursts of high intensity followed by periods of relative calm. The power-law kernel's ability to model long-memory effects and heavy-tailed decay in event clustering makes it particularly well-suited for capturing the complex nature of market activity.

In contrast, \textbf{the exponential kernel, while providing a simpler and more tractable framework, shows limitations in fully representing the observed data}. The buy intensity (blue) under the exponential kernel exhibits fewer and less pronounced peaks, while the sell intensity (yellow) appears more uniform with occasional spikes. This suggests that the exponential kernel's rapid decay rate may oversimplify the temporal dependencies in the data, failing to account for the prolonged influence of past events that is often observed in HFT environments. The power-law kernel's slower decay and ability to model long-range dependencies make it a more robust choice for accurately reflecting the intricate patterns of market behavior.

The superior performance of the power-law kernel can be attributed to its flexibility in modeling real-world phenomena, where events are often influenced by a combination of recent and distant past activities. This is particularly relevant in HFT, where the impact of past trades can persist over varying time scales. The exponential kernel, with its fixed and rapid decay rate, struggles to capture these extended temporal dependencies, leading to a less accurate representation of the data.

\subsection*{Analysing Power-Law Kernel}
\textbf{Simulated Data (Power-Law Kernel)}
\\
The simulated data shows intensity functions that are smoother and more regular compared to the real data. The buy intensity (blue) and sell intensity (orange) exhibit well-defined peaks and decays, reflecting the controlled nature of the simulation. Key observations include:

\textbf{Symmetric Patterns:} The buy and sell intensities appear more balanced, with peaks occurring at roughly similar intervals. This symmetry suggests that the simulation assumes a more idealized market environment, where buy and sell activities are evenly distributed and influenced by the same underlying mechanisms.

\textbf{Controlled Decay:} The decay of intensity after each peak follows a predictable pattern, consistent with the power-law kernel’s ability to model long-memory effects. However, the decay in the simulated data is less erratic than in real data, highlighting the absence of real-world noise and external market influences.

\textbf{Artificial Clustering:} The peaks in the simulated data are evenly spaced, indicating that the simulation likely incorporates a predefined clustering mechanism. While this provides a clean representation of self-exciting processes, it lacks the irregularity and unpredictability seen in real trading environments.

\textbf{Real Data (Power-Law Kernel)}
\\~\\
The real data, in contrast, reveals a more complex and nuanced picture of market activity. The buy and sell intensities exhibit irregular patterns, reflecting the chaotic and dynamic nature of real-world trading. Key observations include:

\textbf{Asymmetric Peaks:} The buy intensity (blue) shows sharper and more frequent spikes compared to the sell intensity (orange). This asymmetry suggests that buy-side activity in real markets is more reactive and clustered, potentially driven by aggressive order placement or market-making strategies.

\textbf{Irregular Decay:} The decay of intensity in the real data is less uniform, with some events showing prolonged influence while others dissipate quickly. This irregularity reflects the varying impact of past trades, influenced by factors such as liquidity, market sentiment, and external news.

\textbf{Noise and Outliers:} The real data includes small fluctuations and outliers that are absent in the simulated data. These irregularities highlight the presence of market noise, such as random order placements or reactions to external events, which are difficult to replicate in simulations.

\textbf{Clustering Dynamics:} While both datasets exhibit clustering, the real data shows more varied cluster sizes and durations. Some clusters are short and intense, while others are prolonged but less pronounced. This variability underscores the complexity of real-world trading, where clustering is influenced by a multitude of factors beyond simple self-excitation.

\newpage
\subsection*{Strategy: Clustering-Based Liquidity Provision}
\normalsize{\textbf{\textsc{Core Idea}}\par }
\normalsize
The strategy leverages the clustering behavior of buy and sell orders, as identified by the power-law kernel, to provide liquidity during periods of high activity. By anticipating when and where clusters of orders are likely to occur, the strategy aims to place orders that capture spreads or benefit from short-term price movements.

\normalsize{\textbf{\textsc{Key Components}}\par }
\normalsize
\begin{enumerate}
    \item \textbf{Identify Clusters}:  
    Use the power-law kernel to detect clusters of buy and sell orders in real-time. The intensity functions can serve as signals for when activity is likely to spike, indicating periods of high liquidity demand.  
    \begin{itemize}
        \item \textbf{Buy Clusters}: Sharp peaks in buy intensity suggest aggressive buying pressure, potentially leading to short-term price increases.  
        \item \textbf{Sell Clusters}: Peaks in sell intensity indicate selling pressure, which may lead to short-term price declines.  
    \end{itemize}

    \item \textbf{Liquidity Provision}:  
    During detected clusters, place limit orders slightly ahead of the anticipated price movement to capture spreads.  
    \begin{itemize}
        \item \textbf{Buy Clusters}: Place sell limit orders at slightly higher prices to profit from the upward price movement.  
        \item \textbf{Sell Clusters}: Place buy limit orders at slightly lower prices to benefit from the downward price movement.  
    \end{itemize}

    \item \textbf{Adjustments}:  
    Continuously monitor the decay of intensity after each cluster. As the influence of past events diminishes, adjust or cancel orders to avoid adverse selection. The power-law kernel’s ability to model long-memory effects helps in predicting how long the impact of a cluster will last.  

    \item \textbf{Risk Management}:  
    \begin{itemize}
        \item \textbf{Order Size}: Adjust order sizes based on the intensity of the cluster. Higher intensity clusters may warrant larger orders, but this should be balanced against the risk of overexposure.  
        \item \textbf{Stop-Loss}: Implement stop-loss mechanisms to limit losses in case the market moves against the anticipated direction.  
    \end{itemize}
\end{enumerate}

\subsubsection*{Advantages}
\begin{itemize}
    \item \textbf{Exploits Market Microstructure}: The strategy takes advantage of the self-exciting nature of HFT data, which is often overlooked by simpler models.  
    \item \textbf{Adaptive to Market Conditions}: By using the power-law kernel, the strategy can adapt to varying market dynamics, including irregular clustering and prolonged dependencies.  
    \item \textbf{Scalability}: The strategy can be applied across multiple assets, as the power-law kernel’s flexibility allows it to capture asset-specific behaviors.  
\end{itemize}

\subsubsection*{Challenges}
\begin{itemize}
    \item \textbf{Latency}: Real-time detection of clusters and placement of orders require ultra-low latency infrastructure, which may be costly to implement.  
    \item \textbf{Noise and Outliers}: The presence of noise in real data can lead to false signals, requiring robust filtering mechanisms.  
    \item \textbf{Competition}: In highly competitive HFT environments, other market participants may also exploit similar patterns, reducing the profitability of the strategy.  
\end{itemize}

\end{document}